# Overcoming the Bradyon–Tachyon Barrier

*Luca Nanni

*corresponding author e-mail: luca.nanni@student.unife.com

Abstract

In this study, the problem of overcoming the infinite energy barrier separating the bradyonic and tachyonic realms is investigated. Making use of the Majorana equation for particles with arbitrary spin and the Heisenberg uncertainty principle, it is proved that, under certain conditions of spatial confinement, quantum fluctuations allow particles with very small mass and velocity close to the speed of light to pass in the tachyonic realm, avoiding the problem of the infinite barrier (bradyon–tachyon tunnelling). This theoretical approach allows an avoidance of the difficulties encountered in quantum field theory when it is extended to particles with imaginary rest mass.

## 1 Introduction

Faster than light particles (tachyons) have been a topic that the most eminent theoretical physicists of the last century, such as Sommerfeld [1], Pauli [2] and Feynman [3], have dealt with in the course of their research. Although there is still no experimental evidence for the existence of tachyons (or even experimental evidence that definitively proves their false physical reality), several researchers have been working to introduce this hypothetical particle into the modern quantum theory without compromising the foundations [4–6]. In particular, Surdashan, Binaliuk and Recami [7–9] have proved that, by introducing the reinterpretation principle and extending the theory of relativity to superluminal frames and objects, it is possible to solve the problem concerning the violation of the principle of causality that since the beginning has led physicists to consider the tachyon as a particle that is incompatible with physical reality [11–12]. In quantum field theory, the tachyon was introduced as a field with imaginary rest mass, but immediately showed conceptual limits that were not easy to solve [13–14]. For instance, spinless particles can be quantised by the Fermi-Dirac statistics but not by those of Bose, as instead happens for spinless bradyons [4]. Moreover, the concept of imaginary rest mass leads to the problem of tachyon vacuum instability, which can be solved by introducing a preferred reference frame [14], which, however, *undermines* one of the pillars of the theory of relativity. Surdashan considers tachyons as a new kind of particle of class III (bradyons are subluminal particles of class I and luxons are luminal particles of class II) travelling always at superluminal speed [5]. The two realms, tachyonic and bradyonic, are separated by an infinite energy barrier placed at $v = c$ [15], which prevents tachyons from slowing down the speed of light and bradyons from overcoming this velocity. However, it cannot be excluded *a priori* that particles travelling at velocities close to the speed of light, like the neutrino, or highly energetic ones, like the quarks, may pass in the tachyonic regime. The quantum fluctuations often have been very useful to explain physical phenomena that are apparently inconsistent with our perception of reality, like the spontaneous radioactive decay or the fusion ignition of a star (as well as explaining phenomena that still have not been proven, like the Hawking Effect). The purpose of this study is to investigate the possibility that, under certain extreme conditions, a massive particle with internal degrees of freedom (spin) may pass from the bradyonic to the tachyonic realm, overcoming, by quantum tunnelling, the energetic barrier separating them. It will be the Heisenberg uncertainty principle that prevents *seeing* the particle crossing the barrier. To the best of my knowledge, there is only a single research article in the literature that attempts to explain the infinite energy barrier [15]. According to this study, using the virial theorem, it is proved that closed systems of massive bradyonic particles that interact electromagnetically originate a sort of *binding energy* that at $v = c$ tends to $-\infty$, cancelling the infinite barrier separating the bradyonic and tachyonic realms. This idea has the advantage of avoiding the problems encountered when addressing the topic in the ambit of the quantum field theory, but is of a semi-classical nature and requires that the particles interact electromagnetically.

In the present research, a new approach based on the relativistic equation of Majorana for particles with arbitrary spin [16], which describes both bradyons and tachyons, is proposed. In particular, it will make use of the *excited states*, defined by Majorana as *unphysical*, that become accessible only at very high energies and that are characterised by a decreasing mass spectrum [16–17]. It is proved that, under certain constraints (spatial confinement and velocities close to that of light), the quantum fluctuations induced by the uncertainty principle may bring the bradyon into the tachyonic realm. This theory could explain from the perspective of quantum mechanics the decay of a massive particle in luxons, bradyons and tachyons; the kinematics of this kind of phenomenon has been studied by Lemke [18] and never reconsidered by other authors. In order to consider the tachyon as a particle that effectively exists, the study of similar decays becomes an important topic for theoretical physics and, in this case, the problem of overcoming the tachyon-bradyon barrier can no longer be ignored.

## 2 Bradyons and Tachyons from the Perspective of the Majorana Equation

The Majorana equation for particles with arbitrary spin [16] is a relativistic equation describing the quantum states of a free particle with rest mass $m_0$ and spin $s$. The time-like solutions, with increasing total angular quantum number $J_n = (s + n)$ and $n = 0,1,2,...$, are infinite component wave functions with positive energies and transform according to the infinitesimal elements of the unitary Lorentz group. These requirements lead to a discrete bradyonic mass spectrum (bradyonic tower) depending on the quantum number $J$:

$$m(J_n) = \frac{m_0(s)}{J_n + 1/2}. \qquad (1)$$

Here, $m_0(s)$ is the rest mass of a particle with spin $s$. Therefore, the particle is in the fundamental state when the reference frame is that of the center of mass. Moreover, the non-trivial terms of the infinite components wave function are $(2s + 1)$: they are proportional to $(v/c)$ and reduce to that of Schrödinger for slow motion. All other states with decreasing mass, defined by Majorana as *unphysical*, have $(2J_n + 1)$ non-trivial components proportional to $(v/c)^n$. It is clear that when the speed of the particle is close to that of light, these components do not become more negligible and, therefore, there are no physical reasons to reject a priori these solutions. Majorana himself states that such states could be accessible only under extreme conditions of high energy; this hypothesis is the first basic assumption of this study. These states will be referred to from now on as *excited states*.

The general solution of the Majorana equation is given by the linear combination of all the infinite states of the bradyonic tower:

$$|\Psi\rangle = \sum_{n=0}^{\infty} c_n |\varphi(J_n)\rangle \quad n = 0,1,2,... \qquad (2)$$

The coefficients $c_n$ are [19]:

$$c_n = \sqrt{\left(\frac{v}{c}\right)^n - \left(\frac{v}{c}\right)^{n+1}}, \qquad (3)$$

by which is obtained the probability $P_n = c_n^* c_n$ that the nth state is occupied. When $n = 0$ (fundamental state), the occupation probability is $(1 - v/c)$ and decreases with the increase of the particle velocity. When $n > 0$, the occupation probability increases with the particle speed up to a maximum, beyond which it starts to decrease fast to zero. An example of the probability trend vs the degree of the excited state is shown in Fig. 1:

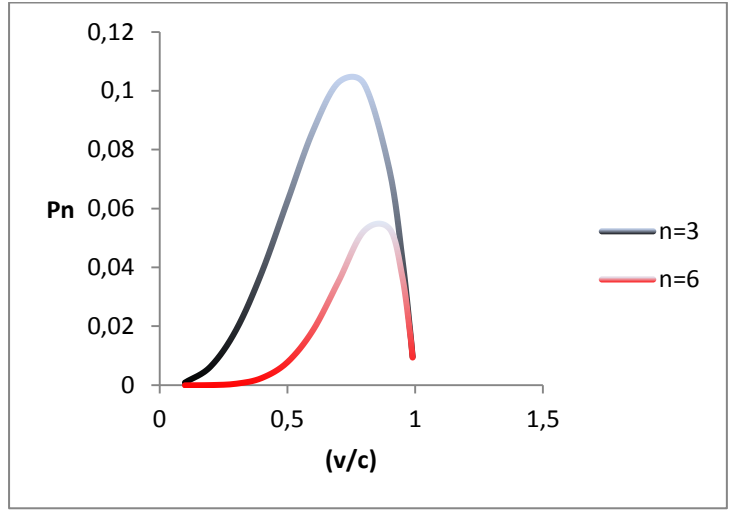

**Figure 1**

It must be noted that by increasing the degree of the excited state, the value of the probability corresponding to the maximum of the curve becomes smaller and the position of this maximum moves to the upper limit of the ratio $v/c$. This behaviour suggests that when the velocity of the particle approaches that of light, the excited states cannot be ignored, even if their occupation probabilities are small. This further strengthens the possible existence of these highly energetic states.

Moreover, it must be noted that when $v = c$, the occupation probability is zero for all quantum states: the light barrier is a region that can never be occupied by massive bradyons.

The Majorana equation also admits space-like solutions satisfying the impulse–energy relationship $E = \pm\sqrt{p^2c^2 - m^2c^4}$ consistent with particles with imaginary rest mass. In this case, the tachyonic mass spectrum is continuous and the energy of the particle may be both positive and negative. In this respect, tachyons recall the quantum behaviour of bradyons and, in principle, can be created in pairs without the necessity to accelerate bradyonic particles through the light barrier [4]. It should be noted that the mass spectrum of the bradyonic tower tends to become continuous, increasing the order $n$ of the excited state. In fact, according to **(1),** the difference between the masses of two contiguous excited states is:

$$\begin{cases} \Delta m(J_n, J_{n+1}) = \frac{m_0}{n+1} & n = 0,1,2,\dots \quad for\ Fermions \\ \Delta m(J_n, J_{n+1}) = \frac{2m_0}{(2n+1)(2n+3)} & n = 0,1,2,\dots \quad for\ Bosons \end{cases} \quad (4)$$

Since it has been shown that the excited states have non-zero probability to be occupied when the particle speed is close to that of light, we can conclude that the mass spectrum of the bradyonic tower gradually tends to become continuous as the velocity of the particle approaches the speed of light. In this sense, the bradyonic and tachyonic spectra bind between them smoothly. This feature of the two mass spectra is the second basic assumption of this study.

The results discussed in this section are the theoretical basis for addressing the problem of the light barrier separating the tachyonic and bradyonic realms.

## 3    Bradyon–Tachyon Tunnelling

Let us consider a particle with rest mass $m_0$, spin $J_0 = s$ and travelling with a velocity close to but lower than the speed of light. According to the Majorana theory, its relativistic impulse is:

$$p = \gamma K_M m_0 u \quad where \begin{cases} K_M = \frac{1}{n+1} & n = 0,1,2,\dots \quad for\ Fermions \\ K_M = \frac{2}{(2n+1)} & n = 0,1,2,\dots \quad for\ Bosons \end{cases}, \quad (5)$$

where $u$ is the particle velocity, $\gamma$ is the Lorentz factor and $K_M$ is defined as the Majorana factor and takes into account the discrete mass spectrum of all infinite solutions of the Majorana equation. The uncertainty affecting the impulse $p$ is due to the uncertainty by which it is possible to know the particle velocity $u$. But this uncertainty affects also the Lorentz factor and the degree $n$ of the Majorana excited state. Since we are considering a particle traveling at a speed close to that of light, it is reasonable to assume that its uncertainty is small enough not to imply significant changes on $\gamma$ and on $K_M$. Therefore, the uncertainty affecting the impulse may be approximated to:

$$\delta p = \gamma K_M m_0 \delta u. \quad (6)$$

Recalling the Heisenberg uncertainty principle and using the **(6),** we get:

$$\delta u \geq \frac{\hbar}{\gamma K_M m_0 \Delta q}, \quad (7)$$

where $\Delta q$ is the uncertainty affecting the position of the particle. Making explicit the Majorana factor for the fermionic and bosonic towers, we get:

$$\begin{cases} \delta u \geq \frac{(n+1)}{\gamma m_0 \Delta q} \hbar & for\ Fermions \\ \delta u \geq \frac{(2n+1)}{\gamma 2 m_0 \Delta q} \hbar & for\ Bosons \end{cases}. \quad (8)$$

On the basis of what has been discussed in the previous section, we can state that the increase of the Lorentz factor is mitigated by that of the Majorana: excited states with a very high total quantum angular number may approach the speed of light spending less energy than the starting particle in the fundamental state. If $\Delta q$ is very small, i.e. the particle is confined in a very small region of space, the fluctuation of its velocity could be large enough as to bring it into the tachyonic regime:

$$u_{Tachyonic} = u_{Bradyonic} + \frac{\hbar}{\gamma K_M m_0 \Delta q} \;. \tag{9}$$

The **(9)** proves that bradyon–tachyon tunnelling may occur, but the probability that it effectively takes place depends on the **(3)**, i.e. the probability that the Majorana excited state with high $J_n$ is occupied. Since this probability is very small, as discussed in the previous section, it is reasonable to expect that bradyon–tachyon tunnelling is very rare. According to the **(9)**, the only conditions to make a similar process physically possible is to have a bradyon with very small rest mass, moving at almost luminal speed in extremely confined spaces. This helps explain why massive particles travelling at superluminal velocities have not yet been experimentally observed, directly or indirectly.

In this study, we are also interested in calculating the mass of the particle after its passage in the tachyonic realm. To do this, we refer to the theory of Park [20], according to which the tachyon has a real mass but a sign that, being reference frame dependent, also could be negative:

$$m = m' \frac{\sqrt{1-\frac{u'^2}{c^2}}}{\sqrt{1-\frac{u^2}{c^2}}} sgn\left(1 - \frac{\mathbf{u}\cdot\mathbf{v}}{c^2}\right), \tag{10}$$

where $u$ is the velocity of the tachyon in the reference frame $S$, $u'$ is its velocity in the reference frame $S'$ and $\mathbf{v}$ is the relative velocity of the two reference frames. The function $sgn\left(1 - \frac{\mathbf{u}\cdot\mathbf{v}}{c^2}\right)$ determines the sign of the mass depending on the magnitude and direction of the velocity. According to this theory, the proper mass of the tachyon is not an absolute quantity.

Let us consider a subluminal particle with small rest mass $m_0$ and velocity $u$ very close to that of light and confined in a very small region of space so that the quantum fluctuation of its velocity is such as to bring it into the tachyonic realm. Since in the Park theory, it is assumed that the energy is conserved [20], we can equal the relativistic energy in the two realms:

$$\begin{cases} \frac{1}{\sqrt{1-\frac{u^2}{c^2}}}\frac{m_0 c^2}{(n+1)} = \frac{1}{\sqrt{\frac{(u+\delta u)^2}{c^2}-1}} m_t c^2 sgn\left(1-\frac{\mathbf{u}\cdot\mathbf{v}}{c^2}\right) & for\ Fermions \\ \frac{1}{\sqrt{1-\frac{u^2}{c^2}}}\frac{2m_0 c^2}{(2n+1)} = \frac{1}{\sqrt{\frac{(u+\delta u)^2}{c^2}-1}} m_t c^2 sgn\left(1-\frac{\mathbf{u}\cdot\mathbf{v}}{c^2}\right) & for\ Bosons \end{cases}, \tag{11}$$

where $m_t$ is the mass of the tachyon and $(u + \delta u)$ is greater than the speed of light. Reworking the **(11)**, we get:

$$\begin{cases} m_t = \dfrac{m_0}{(n+1)} \sqrt{\dfrac{(u+\delta u)^2 - c^2}{c^2 - u^2}} \, sgn\left(1 - \dfrac{\mathbf{u} \cdot \mathbf{v}}{c^2}\right) & for\ Fermions \\ m_t = \dfrac{2m_0}{(2n+1)} \sqrt{\dfrac{(u+\delta u)^2 - c^2}{c^2 - u^2}} \, sgn\left(1 - \dfrac{\mathbf{u} \cdot \mathbf{v}}{c^2}\right) & for\ Bosons \end{cases} \quad (12)$$

The equations **(12)** show that:

a) The magnitude of the relative velocity $\mathbf{v}$ is equal to $\delta u$; therefore, the sign of the function $sgn\left(1 - \dfrac{\mathbf{u} \cdot \mathbf{v}}{c^2}\right)$ is always positive. We conclude that bradyon–tachyon tunnelling leads the massive particle to a tachyon behaviour with positive energy.

b) The mass of the tachyon may assume any value (continuous mass spectrum) according to the Majorana equation, since the uncertainty $\delta u$ varies continuously. However, the value of the mass holds the memory of the excited states of the bradyon.

c) The velocity of the particle in the tachyonic realm is slightly greater than the speed of light, and its quantum fluctuation can return the particle to the bradyonic realm following an opposite tunnelling process. In this sense, we can think of the particle as being in equilibrium around the light barrier.

## 4 Conclusion

In this study, we have proved that massive particles with very small rest mass, subluminal velocity close to the speed of light and confined in a small region of space may transit in the tachyonic realm, overcoming the light barrier. Moreover, as soon as the particle is in the superluminal regime, it may return to the bradyonic realm following an inverse mechanism of quantum tunnelling. A similar physical system behaves as if it were in equilibrium around the light barrier, which is governed by quantum fluctuations affecting its velocity. The only speculation we made use of in elaborating this theory is the existence of the excited states with decreasing mass spectrum provided by the Majorana equation for a particle with arbitrary spin. The probability of occurrence of the bradyon–tachyon tunnelling depends on the probability of the occupation of an excited state with a very high total angular quantum number. We proved that this last probability is very low even at the velocities approaching that of light. Therefore, it is also expected that bradyon–tachyon tunnelling is a very rare phenomenon.

The results of this study may help to explain the phenomena of decay of a subluminal massive particle in tachyons, luxons and bradyons [18]. This is the subject of a further theoretical study, also based on the Majorana equation, with the purpose of proposing a new way to prove experimentally the existence of tachyons.

**References**

[1] A. Sommerfeld, K. Akad, *Wat. Amsterdam Proc. 8*, 346 (1904)
[2] J.H. Brennan, *Time Travel*, Liewellin Publication, U.S.A (1997)
[3] P.R. Feynman, *Phys. Rev.*, 76, 749 (1949)
[4] G. Feinberg, *Phys. Rev.*, 159(5), 1089 (1967)
[5] O.M.P. Bilaniuk, E.C.G. Surdashan, *Phys. Today*, 22(5), 43 (1969)
[6] O.M.P. Bilaniuk, *Phys. Today*, 22(12), 47 (1969)
[7] M.E. Arons, E.C.G. Surdashan, *Phys. Rev.*, 173(5), 1622 (1968)
[8] O.M.P. Binaliuk, V.K. Despande, E.C.G. Surdashan, *Am. J. Phys.*, 30(2), 718 (1962)
[9] E. Recami, R. Mignani, *Il Nuovo Cimento*, 4(2), 209 (1974)
[10] E. Recami, *Found. Phys.*, 31, 1119 (2001)
[11] W.B. Rolnick, *Phys. Rev.*, 183, 1105 (1969)
[12] R.G. Newton, *Phys. Rev.*, 162, 1274 (1967)
[13] S. Hamamoto, *Progress of Theor. Phys.*, 48(3), 1037 (1971)
[14] V.F. Perepelitsa, *arxiv:1407.3245v4 [Physics.gen-ph]*, (2015)
[15] M. Ibison, *arxiv:0704.3277 [Physics.gen-ph]*, (2007)
[16] E. Majorana, *Il Nuovo Cimento*, 9, 335 (1932) – English Translation by C.A. Orzalesi in *Technical Report*, 792, University of Meryland (1968)
[17] D.M. Fradkin, *Am. J. Phys.*, 34, 314 (1966)
[18] H. Lemke, *Phys. Rev. D*, 22(6), 1342 (1980)
[19] L. Nanni, *arxiv:1603.05965 [Physics.gen-ph]*, (2016)
[20] M.I.Park, Y.J. Park, *Il Nuovo Cimento*, 111B(11), 1333 (1966)